\documentclass[12pt,preprint]{aastex}

\newcommand{\feh} {\mbox{\rm [Fe/H]}}
\newcommand{\teff} {\mbox{\rm $T_{eff}$}}

\begin{document}


\title{The CH(G) index as a new criterion for selecting red giant stars}
\author{Y.Q. Chen, G. Zhao,  K. Carrell, J.K. Zhao and K.F. Tan}

\altaffiltext{}{Key Laboratory of Optical Astronomy, National Astronomical Observatories, Chinese Academy of Sciences, Beijing, 100012, China; cyq@bao.ac.cn.}

\begin{abstract}
We have measured the CH G band (CH(G)) index for evolved stars in the
globular cluster M3 based on the SDSS spectroscopic survey. It is found that
there is a useful way to select red giant branch (RGB) stars
from the contamination of other evolved stars, such as
asymptotic giant branch (AGB) and red horizontal branch (RHB) stars, by using the CH(G) index
versus $(g-r)_0$ diagram if the metallicity is known from the spectra.
When this diagram is applied to field giant stars with similar metallicity 
we establish a calibration of $CH(G) =1.625(g-r)_0-1.174(g-r)_0^2-0.934$.

This method is confirmed by stars with $\feh \sim -2.3$
where spectra of member stars in globular clusters
M15 and M92 are available in the SDSS database.
We thus extend this kind of calibration to every individual metallicity 
bin ranging from $\feh \sim -3.0$ to $\feh \sim 0.0$
by using field red giant stars with $0.4\,\leq\,(g-r)_0\,\leq\,1.0$.
The metallicity-dependent calibrations give,
$CH(G)=1.625(g-r)_0-1.174(g-r)_0^2+0.060\feh-0.830$ for $-3.0 <\feh \leq -1.2$ and
$CH(G)=0.953(g-r)_0-0.655(g-r)_0^2+0.060\feh-0.650$ for $-1.2 <\feh <0.0$.
The calibrations are valid for the SDSS spectroscopic data set, and
they can not be applied blindly to other data sets.
With the two calibrations, a significant
number of the contaminating stars 
(AGB and RHB stars), were excluded and 
thus a clear sample of red giant stars is obtained by selecting stars
within $\pm 0.05\,mag$ of the calibration.
The sample is published online and it is expected
that this large and clean sample of RGB stars
will provide new information on the formation and evolution of the Galaxy.

\end{abstract}


\section{Introduction}
Late type stars constitute the most important tracers for
understanding the chemical and kinematical evolution of the Galaxy.
Among them, unevolved stars, dwarf and subgiant stars, are 
widely used to probe different stellar populations
of the Galaxy in the solar neighborhood. However, evolved
stars with bright absolute magnitude
are necessary targets to extend the Galactic study far from the 
solar neighborhood where the Galactic halo is the main population.
In this regard, horizontal branch stars with a constant luminosity
are the most common stellar tracers and thus are widely used in 
Galactic halo study. However, we note that the most populated targets of evolved
stars in the halo are red giant stars, which  will
constitute a larger sample of targets for statistical study. In this sense,
red giant stars are important stellar
tracers for Galactic evolution.

The SDSS and its extensions provide $ugriz$ photometry and low resolution
spectra for a large amount of Galactic stars, including a huge amount of red giant stars having the main spectral
types of G and K (hereafter GK red giants)\citep{Abazajian09,Yanny09,Aihara11,Ahn12}.
There are some works on the Galaxy based on unevolved stars
(main-sequence and turnoff stars) of the SDSS spectroscopic survey \citep{Allende06,Carollo10,Bond10}
and a few works adopt later type M giants as stellar tracers \citep{Palladino12}.
As we stated above, unevolved stars reach a smaller distance range
than that of evolved stars, and later type M giants mainly
represent the metal-rich or young populations \citep{Palladino12}.
Instead, GK red giant stars
have the advantage of tracing the metal-poor and old populations
of the Galaxy extending to a distance of at least 20 kpc.
However, there is no work on Galactic evolution 
using GK red giant stars as stellar tracers based on the SDSS survey
in the literature.
Thus, it is of high interest to investigate the chemical and kinematic properties of
different populations in the Galaxy by using GK red giant stars from
the SDSS spectroscopic survey, which will be the scope of our next paper.
It is expected
that a large sample of GK red giant stars in the SDSS survey
will provide new information on the formation of
the halo, the division of the inner and outer halo
and the transition from the halo to disk.

Before a statistical study
can be carried out, clear selection of red giant stars without
contamination of stars from other stages is important.
It is known that evolved stars with GK spectral type in the SDSS spectroscopic survey span the entire
RGB mixing with the red clump, RHB and early-AGB phases.
The contamination of these stars in the RGB sample could
affect the spatial, kinematic and chemical distributions
in the statistical study. In particular, distances of red 
giant stars beyond the solar neighborhood
are usually estimated
from interpreting the fiducial sequence (hereafter FS) of globular clusters (GCs) or
isochrones of theoretical models.
In this way, wrong distances will be provided for
the contaminating stars (such as red clump, RHB and
early-AGB stars) if they are assumed to be RGB stars
since they have similar colors but
very different luminosities at a given metallicity.
In principle, stellar parameters (temperature, gravity and metallicity)
from the SSPP \citep{Lee08a,Lee08b} in the SDSS database
can be used to separate evolved and unevolved stars,
with $logg < 3.5$ being the former. However, 
since $logg$ values provided by the SSPP pipeline have quite large
uncertainties, they cannot be used to further classify 
among evolved stars.
In this work, we are searching for
the possible criterion for separating the early-AGB and RHB stars
from the RGB by using the line index measured from member stars
of globular clusters and field stars in the SDSS survey.

\section{Data and Line Index Measurements}
There are eight GCs in the SDSS spectroscopic survey that
have enough spectra of member stars for statistical study, and
\citet{Smolinski11a} have presented these cluster stars.
For field stars, 
we select a sample of stars from the SDSS DR9 catalog with a $(g-r)_0$ range
of $-$0.2 to 1.0 $mag$, $\feh$ from $<-2.8$ to $0.0$ and $logg$ less
than 3.5 dex, where the main
targets are giant stars. 
The sample is limited to stars
with spectra available in the DR7 database, but the stellar
parameters and $(g-r)_0$ were taken from the SDSS DR9 database.
This limitation of using spectra in DR7 (instead of DR9)
will reduce the star number in the field sample, but
we assume that there are enough stars in the DR7 database
for high statistics.

We have measured the CH(G) index (CH G band at $\lambda4300$\AA) 
defined by \citet{Lee99} and the spectral index S(3839) (CN band at $\lambda3883$\AA\ defined by \citet{Norris81}, following
the same definitions adopted in \citet{Smolinski11b}.
Note that the released spectra of the SDSS DR7 database are flux calibrated
and the wavelength is shifted such that measured velocities will be
relative to the solar system barycenter at the mid-point of each 15-minute exposure.
We thus corrected the wavelength with the redshift of $elodierv_z = (elodierv-7.3)/c$
where $elodierv$ is taken from the sppParams table of the SDSS DR7 database
(in order to fit our spectra from DR7; and we checked that
they are generally in agreement with DR9), 
$c$ is the speed of light, and 7.3 km~s$^{-1}$ is an empirically 
derived offset putting the $elodierv$ of all stars on a system consistent 
with that of other literature measures of known radial velocity standards as
described in \citet{Lee08b}. Then we interpolate the spectra
and produce new spectra with 5,000 points for the wavelength range of 3840-4500\AA\ (versus $\sim$700 points
in the original spectra) so that the measurements
of the line index will not depend on the chosen points at the edge of the defined bands
and thus the line index will become
more consistent among stars.
 We have checked that both sets of data generally agree quite
well. Errors are less than 0.03 mag for $(g-r)_0$ 
based on $g$ and $r$ SDSS photometric errors,
less than 0.015 $mag$ for the S(3839) index and less than 0.012 $mag$ for the CH(G) index.
To determine the errors on the line index we included errors in the flux (available 
in the SDSS spectra fits files) and in the wavelength shift (via the radial velocity errors) 
since these are the two dominant sources of uncertainty in our determination.  The comparison 
of our indices with those of Smolinski et al. (2011b) shows a systematically higher value 
for the CH(G) index and lower value for the S(3839) index by the order of $0.02\,mag$ 
in our work with a small scatter of $0.003\,mag$.  
The systematic deviation disappears if we adopt the redshift available in the fits file headers.  
Therefore, the small scatter in the line index comparison indicates a good agreement 
between Smolinski et al. (2011b) and our work. The wavelength correction by using
$elodierv$ in our work is more reasonable than using the redshift in the fits file headers
since this is the way that the SDSS survey presents its spectra. But both works have
internally consistent values for stars and thus this difference will not significantly affect the results.

\section{Results}
\subsection{The $(g-r)_0$ versus CH(G) diagram in M3}
Among the eight GCs in the SDSS spectroscopic survey, only M3 has
a significant number of AGB stars, while others do not have
many early-AGB and RHB member stars. Thus we start the study
from globular cluster M3 in this work. We adopted the member star
list of M3 from \citet{Smolinski11b} (their Table 3) and obtained photometry and
stellar parameters from the SDSS DR9 database \citep{Ahn12}. With the adopted distance modulus of 15.070
in $V$ band and reddening of 0.010 from the new version of
\citet{Harris96} published in 2010\footnote{http://physwww.mcmaster.ca/∼harris/mwgc.dat}
we calculated
the absolute magnitude in $g$ band ($M_g=g_0-15.070-3.1*0.010$) for
each star.

There are 77 member stars in the list of \citet{Smolinski11b}, but
we exclude four BHB stars with $(g-r)_0 < 0.0$ since they are
not in the color range of GK red giant stars. Moreover, we
found one star with the identification numbers 
(plate-mjd-fiberid) of 2475-53845-486 is
significantly outside the FS of M3, which was excluded as well. 
Among the remaining 72 member stars, three groups are divided 
according to their evolutionary stages. There are 
12 RHB (open squares), 10 early-AGB (open diamonds) and 50 RGB stars
(filled circles), which are shown in Fig.~1.
The solid line in the upper-left panel of Fig.~1 shows the observed
FS of M3 based on \citet{An08} and the dash-dotted line shows a
theoretical isochrone of 11.5 Gyr, which is close to 11.3 Gyr
for M3. Also shown are dotted lines for 8 and 
14 Gyr isochrones at $\feh = -1.5$ from the Dartmouth isochrones \citep{Dotter08}.
It seems that
age variation from 8 to 14 Gyr has no significant effect on the CMD
and thus it is possible to extrapolate our result to field stars
with different ages in the following sections.

The upper-right panel of Fig.~1 shows the CH(G) index
as a function of $(g-r)_0$ with three groups of stars indicated
by different symbols. The lower two panels present the S(3839) index
as functions of $(g-r)_0$ and $M_g$, and stars with enhanced S(3839) index
are indicated by additional pluses in Fig.~1.
 It is interesting that the RGB is clearly
separated from early-AGB stars in the $(g-r)_0$ versus
the CH(G) index diagram despite their similar colors.
In the S(3839) vs. $(g-r)_0$ diagram, RGB stars
with $(g-r)_0>0.4\,mag$ show two branches with a clear gap in between.
Early-AGB stars lie in between but are
located closer to the lower branch.
There is a hint of lower CH(G) indices
for S(3839)-strong RGB stars as compared with S(3839)-weak RGB stars, but
it is difficult to separate them considering significant
scatter in the CH(G) index at a given $(g-r)_0$ color.
With significantly lower $(g-r)_0$ colors, RHB stars have
even lower CH(G) indices due to their higher temperatures.
The implication from Fig.~1 is that the CH(G) index is a better 
criterion than the S(3839) index for excluding early-AGB stars from RGB stars.
Moreover, in comparison with the $\teff$ versus $logg$ diagram in Fig.~2, we can see that
the CH(G) index is a more successful way to exclude early-AGB and RHB/BHB
stars.

The origin for the lower CH(G) indices for AGB stars than those of RGB stars at similar colors
is the so-called  weak G-band effect first noted by \cite{Zinn73}.
This effect comes from a combination of low carbon abundances and low gravities 
in the AGB stars \citep{Suntzeff81,Briley90}. The lower carbon abundance in AGB stars than
RGB stars at similar colors was theoretically predicted since \cite{Sweigart79} and
very recently by \cite{Angelou11}, who
proposed a theoretical model of carbon evolution for M3 and
showed that carbon abundances 
decrease after the bump of the RGB due to the onset of extra mixing.
With bright luminosities, the  early-AGB stars of M3 in our work 
have lower gravities than RGB stars at similar colors and are in 
a later stage after the bump of the RGB.
Both factors contribute to the significant gap between the early-AGB stars and lower RGB stars
in the $(g-r)_0$ versus CH(G) diagram.

\begin{figure*}[ht]
\epsscale{1.0} \plotone{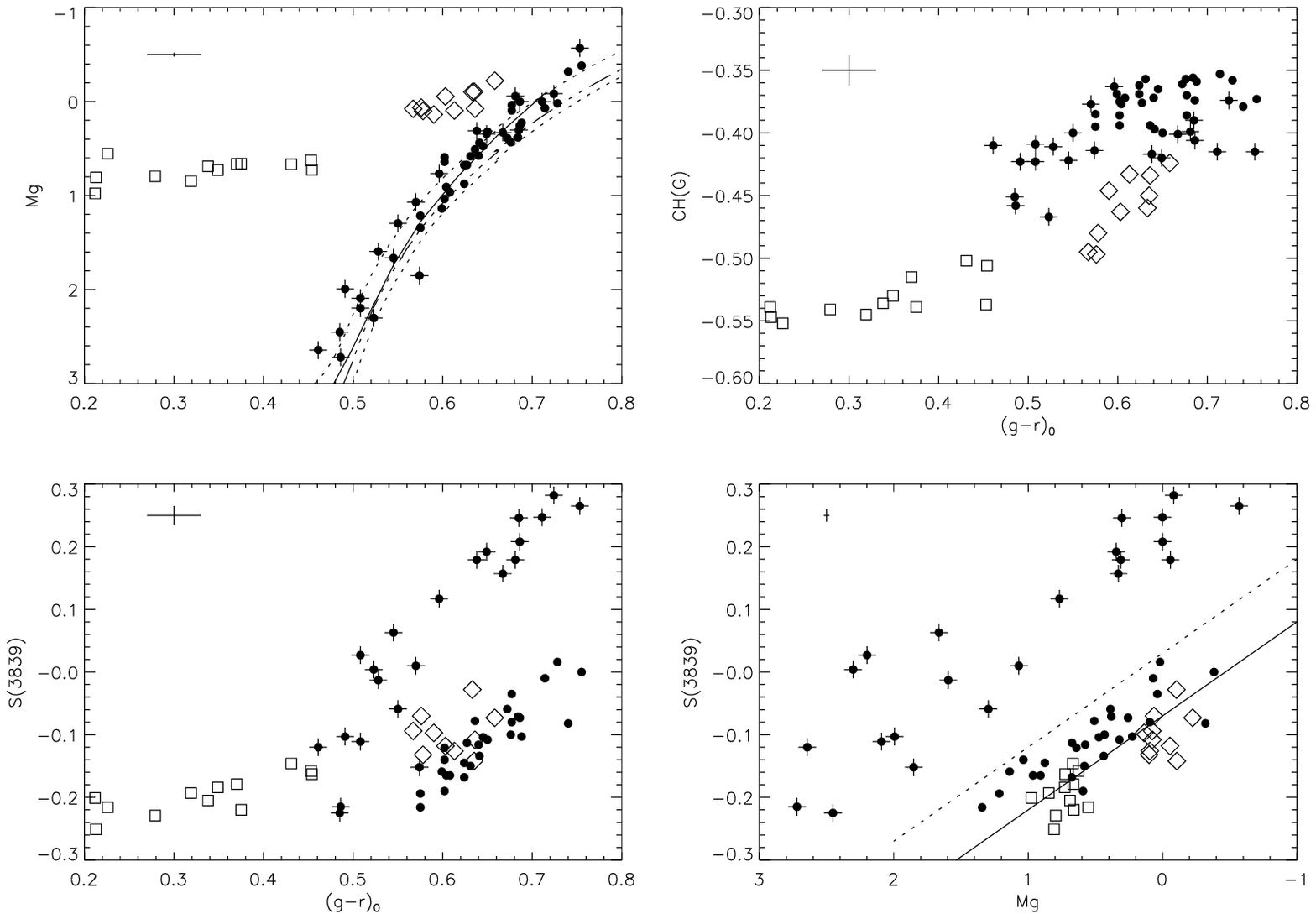}
\caption{Upper panels: The absolute magnitude $M_g$ and the CH(G) index
as a function of $(g-r)_0$ for RHB (squares), early-AGB (diamonds),
and RGB (filled circles) stars in M3. The solid line
is the observed FS of \citet{An08} and the dashed lines show 
theoretical isochrones of 8, 11.5 and 14 Gyr at $\feh = -1.5$ from \citet{Dotter08}.
Lower panels: The S(3839) indices are
shown as functions of  $(g-r)_0$ and the absolute magnitude $M_g$ for stars
in M3. The solid line shows the fit to S(3839)-normal stars and the 
S(3839)-enhanced stars
above the dashed line are plotted with additional pluses in all panels.}
\label{fig1}
\end{figure*}

\begin{figure}[ht]
\epsscale{1.0} \plotone{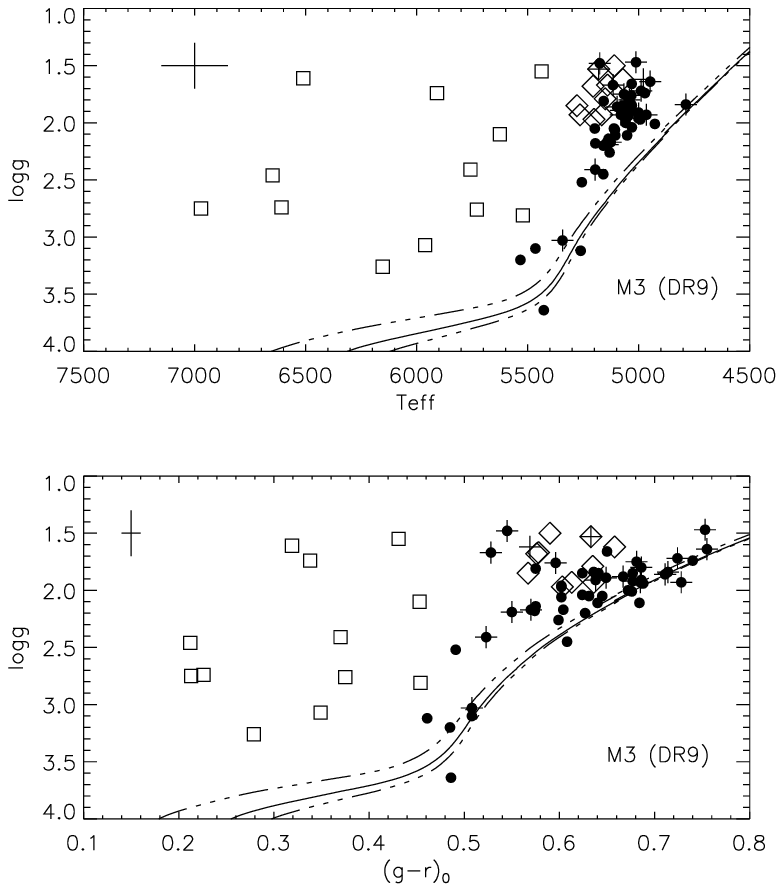}
\caption{The $\teff$ versus $logg$ diagram
for member stars in M3. The symbols are the same as in Fig.~1.}
 \label{fig2}
\end{figure}

In order to check if the CH(G) index can successfully separate
RGB stars from other contaminations at a given metallicity similar
to M3, we select a sample of stars from SDSS DR9 with $0.1\,\leq\,(g-r)_0\,\leq\,0.8$
$mag$, $-1.6\,\leq\,\feh\,\leq\,-1.4$ dex and $logg< 3.5$ 
dex, where the main
targets are giants.
In order to avoid the early type stars, we limit the stars to have
SSPP temperatures from 3000\,K to 10\,000\,K and the signal-to-noise ratio larger than 10.
We do not limit the low edge of $logg$, but we note that our selected stars with
SSPP parameters available
in the SDSS database have surface gravity of $logg>0.0$.
Fig.~3 shows the $(g-r)_0$ versus CH(G) index for the above selected
field stars with locations of member stars of M3 overplotted.
It seems 
that most stars with $(g-r)_0>0.4\,mag$ in our selected field sample
are RGB stars and they match the locations of RGB stars of M3. 
In order to apply the method for further use,
 we have established a calibration between the CH(G) and $(g-r)_0$
for RGB stars of $CH(G)=1.625(g-r)_0-1.174(g-r)_0^2-0.934$.
The dashed lines follow the above calibrated line
with a deviation of $0.05$ dex in the CH(G) index at a given $(g-r)_0$ color.
Stars within the dashed lines are classified as RGB stars.
Stars significantly lower than the calibration may consist 
of early-AGB and RHB stars as indicated by member stars in M3. 
Turnoff and main sequence stars are clumped at the blue edge
with $(g-r)_0$  from 0.2 to 0.4 $mag$, they have CH(G) index lower
than $-0.50$, and overlap with HB stars. 
RHB stars have a wider range of $(g-r)_0$ from 0.2 to $0.5 mag$ but
also have CH(G) indices
less than $-0.50$.

\begin{figure}[ht]
\epsscale{1.0} \plotone{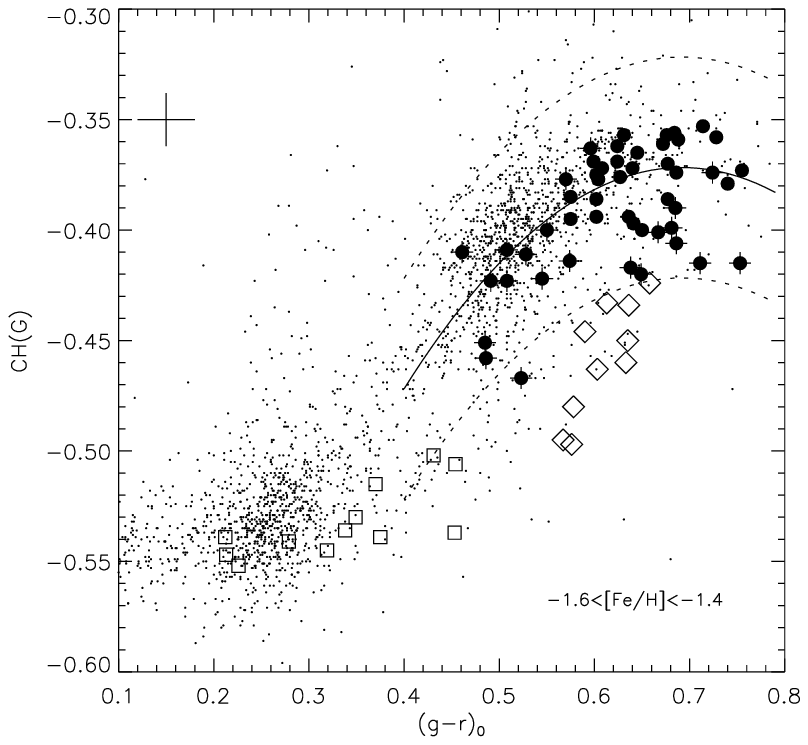} \caption{The $(g-r)_0$ versus CH(G) index
for field stars with $0.1\,\leq\,(g-r)_0\,\leq 0.8\,mag$, $-1.6\,\leq\,\feh\,\leq -1.4$ and $logg<3.5$
in the SDSS DR9 catalog (small dots). Member stars from M3 
are indicated with the same symbols as in Fig.~1. 
The calibration of $CH(G)=1.625(g-r)_0-1.174(g-r)_0^2-0.934$ and the deviations
of $0.05\,mag$ are indicated by solid and dashed lines.
Stars within the dahsed lines are classified as RGB stars.
}
 \label{fig3}
\end{figure}

\subsection{Applying this method to $\feh \sim -2.3$}
In order to extend the study to other metallicities, we investigated
the spectroscopic data of the SDSS survey for M92 and M15,
which have metallicities of $\feh \sim -2.3$. Fig.~4 shows the evolutionary
stages, the CH(G) and S(3839) indices as a function of
 $(g-r)_0$, and the absolute magnitude versus S(3839) index.
The same symbols as in Fig.~1 are used and turnoff stars
from the GCs are shown as crosses.
The same results are found at this metallicity but the scatter is
slightly larger than that in M3. Specifically, the
only possible AGB star is located at the lower edge in the diagram
of the $(g-r)_0$ versus CH(G) index, and subgiants
and turnoff stars generally
have lower CH(G) indices than those of RGB stars.
In particular, a few stars with enhanced S(3839) indices
lie in the middle part of the $(g-r)_0$ versus CH(G) index diagram
among member stars. Thus, the S(3839) index is not taken into account
in selecting RGB stars in the following sections.
Note that the deviation in the CH(G) index is large for one
RGB star with the plate-mjd-fiberid numbers of 1960-53289-530
and the reason for this odd value is unknown.

In connection with possible differences between cluster and field stars, 
there are more cluster stars than field stars at the red end of RGB stars in M3 (see Fig.~3), which
are lacking in M15/M92 (see Fig.~5). The difference in stellar color distribution between cluster and field stars
comes from the target selection of the clusters in the SDSS spectroscopic survey because
stars within the cluster tidal radius are favorably selected as targets and thus the color range of
the targets depends on the distance of the cluster. This selection criterion is not applied in the
target selection of the general field. In addition,
there is a hint that the CH(G) band index for field stars are slightly higher than those of the
cluster stars at a given color in Fig.~5 and possibly in the blue end of RGB stars in Fig.~3. 
This discrepancy comes from the different absolute magnitudes of the stars in the sense that 
bright field stars with $g_0 <16\,mag$
show a slightly higher CH(G) band index than that of fainter stars with $g_0 >19\,mag$ in Fig.~5.
The quality in the spectra of cluster stars is higher than the average value of field stars
at the blue end of RGB stars (and also fainter magnitude), which leads to a slightly lower CH(G) index.
In addition, the uncertainty in reddening for field stars could be larger than that of cluster stars, which
could also contribute to the larger scatter (via $(g-r)_0$ color) for field stars as compared with
cluster stars in this diagram.

\begin{figure*}[ht]
\epsscale{1.0} \plotone{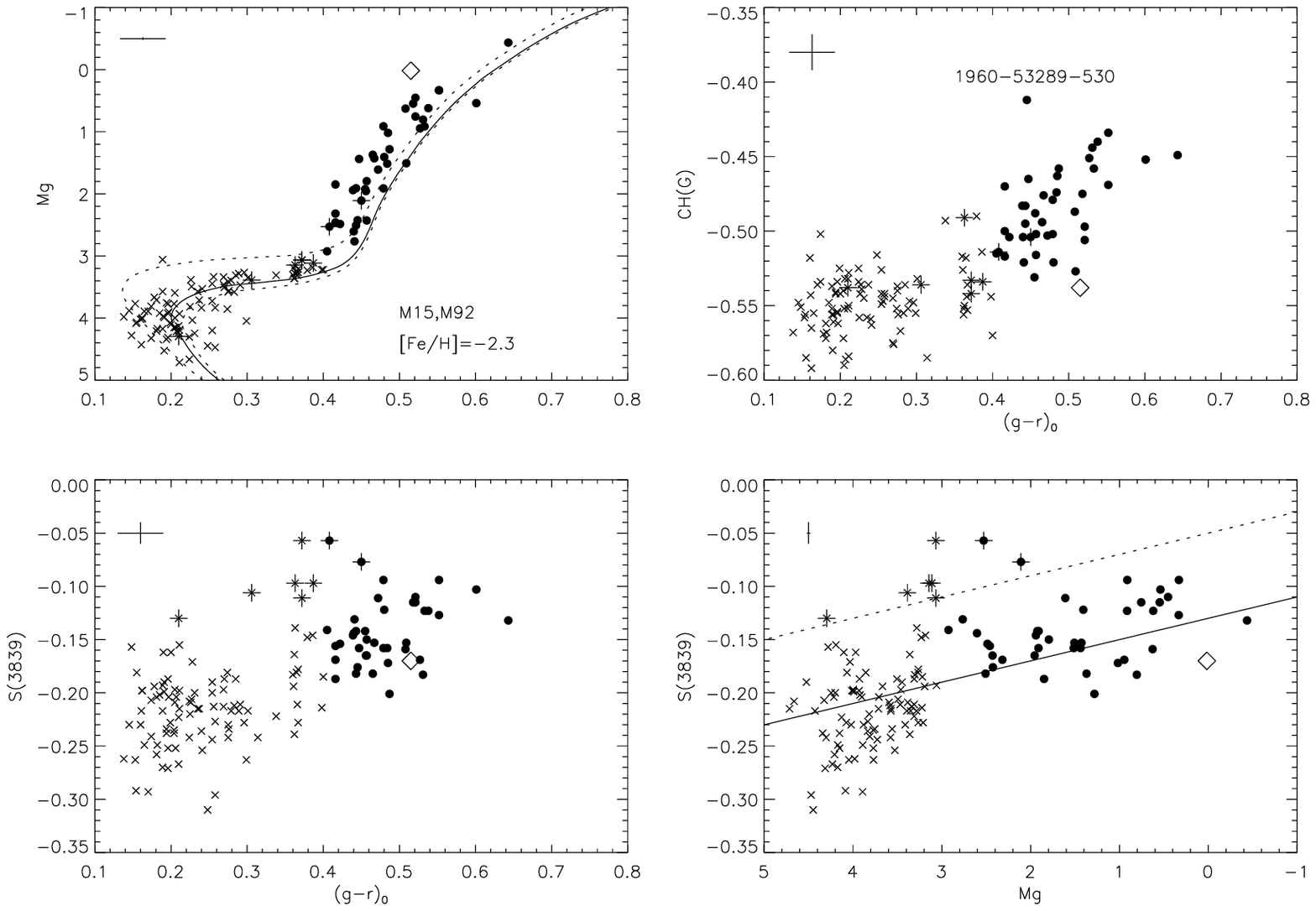} 
\caption{The CH(G) and S(3839) indices versus
  $M_g$ and $(g-r)_0$ for member stars in M92 and M15 with $\feh \sim -2.3$.
Turnoff and subgiant stars are indicated by crosses and other symbols are the same
as in Fig.~1.}
 \label{fig4}
\end{figure*}

In a similar procedure as for M3, we selected a sample of field
stars from the SDSS DR9 database with a $(g-r)_0$ range
of 0.1 to 0.8 $mag$, $\feh$ from $-2.4$ to $-2.2$ dex and $logg$ less
than 3.5 dex, and the $(g-r)_0$ versus CH(G) index
diagram is shown in Fig.~5. The member stars from
 M92 and M15 are overplotted. Then we established a calibration
for the selected field stars located in the RGB region of Fig.~5.
We found that the coefficients 
are very similar to those at $\feh = -1.5$, but the constant is different.
We thus adopted the same $(g-r)_0$ coefficients and obtained the calibration
of $CH(G)=1.625(g-r)_0-1.174(g-r)_0^2-0.982$ for $\feh = -2.3$ as
shown with the solid line in Fig.~5.
Again, the two dashed lines show the locations of RGB stars around the
calibration with a deviation of 0.05 $mag$ in the
CH(G) index at a given $(g-r)_0$ color, and 
stars within the dashed lines are classified as RGB stars.
In the selection, many CEMP stars (see later discussions in Sect. 3.3)
with significant enhancement of the CH(G) index were avoided with the upper
cut of the RGB calibration, and
a few early-AGB stars or AGB-related stars are excluded
with the low cut of the RGB calibration. Finally, it may be more reasonable to
exclude blue stars where the strength of the CH(G)
index starts to decrease for turnoff stars and RHB stars.

\begin{figure*}[ht]
\epsscale{1.0} \plotone{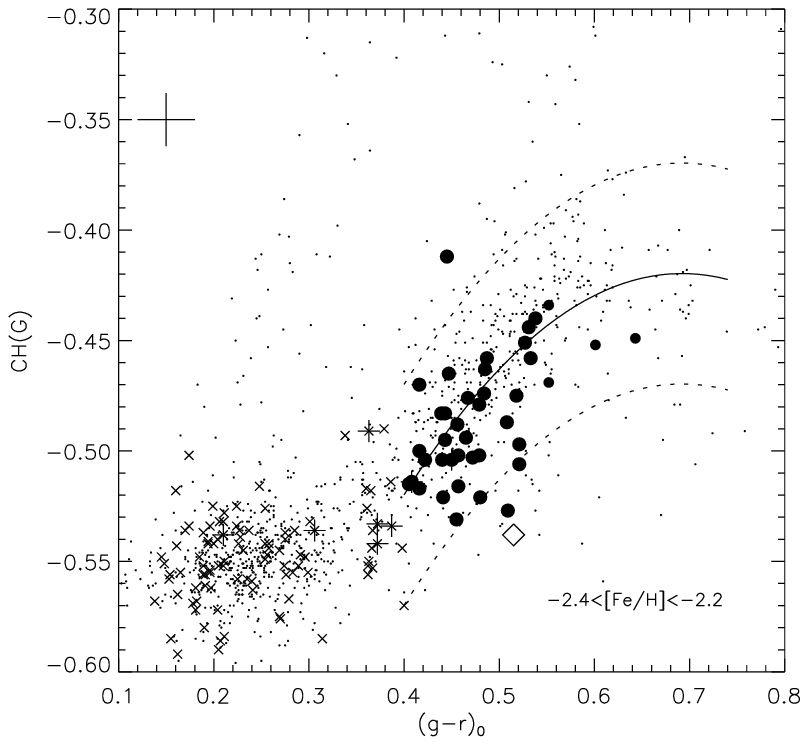} 
\caption{The $(g-r)_0$ versus CH(G) index
diagram for field stars with $\feh = -2.4$ to $-2.2$, $(g-r)_0$
of 0.1 to 0.8 $mag$ and $logg$ less
than 3.5 dex in the SDSS DR9 database. Member stars from M92 and M15 with
$\feh \sim -2.3$ are overplotted. The calibration
of $CH(G)=1.625(g-r)_0-1.174(g-r)_0^2-0.982$ is shown with
the solid line, and stars within the deviations of 0.05 $mag$ (dashed lines)
are classified as RGB stars.
}
 \label{fig5}
\end{figure*}

\subsection{The metallicity-dependent calibration for RGB stars}
In view of similar coefficients but different constants
in the calibrations between $(g-r)_0$ and CH(G) index
at $\feh=-1.5$ and $\feh=-2.3$, we attemp to establish a 
metallicity-dependent constant for the calibration.
For this purpose, we need to extend and obtain
constants for more metallicity bins. First, we selected
field stars with $(g-r)_0$
of $0.1$ to 1.0 $mag$ and $logg$ less
than 3.5 dex for each metallicity bin ranging from $\feh<-2.8$ to
$\feh=0.0$ with a step of 0.2 dex.
We assume that the main populations of field stars, selected based on $(g-r)_0$ 
of 0.4 to 1.0 $mag$ and $logg$ less
than 3.5 dex at a given metallicity, are RGB stars in the SDSS DR9 database.
This assumption is valid for $\feh=-1.5$ and $\feh=-2.3$ as described above.
Again, CEMP stars with significant enhancement of the CH(G) index at
a given color and metallicity were avoided and we excluded blue stars with
$(g-r)_0<0.4$ being turnoff stars or RHB/BHB stars.
Then we used the above-adopted coefficients for $(g-r)_0$ and obtained a set of
different constants for the individual metallicity bins in the fitting of RHB stars.
Then we fit the constants with a metallicity-dependent linear function, and
the final calibration gives
$CH(G)=1.625(g-r)_0-1.174(g-r)_0^2+0.060\feh-0.844$.

Fig.~6 shows the location of selected RGB stars around the calibration
(the solid line for $\feh \leq -1.2$ and long dashed line for $\feh > -1.2$) within 
the deviation of $0.05\,mag$ (the two dashed lines for $\feh \leq -1.2$). 
Note that we only show the ranges of $(g-r)_0=0.1-0.8$ for $\feh \leq -1.2$ and
of $(g-r)_0=0.1-1.0$ for $\feh > -1.2$, corresponding to GK giants.
There are small numbers of stars outside the red cuts until M giants
become significant at the metal-rich end.
We should keep in mind that this calibration is only one possible way to
select most RGB stars from the SDSS database. It is by no means the only
solution to pick out RGB stars. In particular, this calibration 
deviates from the centering points of field giant stars and it gives too
high values for the high metallicity end and perhaps too low values
for the low metallicity end. We have thus provided an additional calibration
of $CH(G)=0.953(g-r)_0-0.655(g-r)_0^2+0.060\feh-0.650$ for $\feh > -1.2$
to fit especially the high metallicity data in a better way. The solid line
and two dashed lines for in Fig.~6 $\feh > -1.2$ is the new calibration and 
its deviation by the order od of $0.05\,mag$.
This new calibration may be more reasonable in view of the fact that
the color range of RGB stars is shifted to the red end as the metallicity increases
and the age span in field stars become very large for $\feh > -0.5$.
Despite the possible deviation between the two calibrations at the high metallicity end, most field giant 
stars can successfully be picked out by selecting stars within a deviation of $0.05\,mag$ in the
CH(G) index from both calibrations. 
A catalog of the selected RGB stars are published electronically and a sample
table consisting of the first ten stars is presented in Table 1.

\begin{table}
\caption{The first ten stars in the selected RGB sample.}
\setlength{\tabcolsep}{0.08cm}
\begin{tabular}{rrrrrr}
\noalign{\smallskip}
\hline
plate & mjd    & fiber & $\feh$ & $(g-r)_0$ &  CH(G) \\
\noalign{\smallskip} \hline
      2669&    54086   &   534&   -3.03&  0.564& -0.452\\
      2724 &   54616  &    324&   -3.15&  0.636& -0.485\\
      2271  &  53726 &       2&   -3.15&  0.460& -0.555\\
      1664   & 52965&      452&   -3.06&  0.441& -0.543\\
      2547    &53917     & 163&   -3.04 & 0.456& -0.503\\
      2534&    53917    &  546&   -3.26&  0.622& -0.422\\
      2553 &   54631   &   439&   -3.00 & 0.446& -0.500\\
      2186  &  54327  &    277&   -3.06&  0.468& -0.540\\
      2302   & 53709 &     182 &  -3.55&  0.544& -0.509\\
      2300    &53682&      472  & -3.01&  0.607& -0.417\\
\hline
\noalign{\smallskip}
\end{tabular}
\end{table}

Note that there is an interesting group of stars with significantly
higher  CH(G) index mainly at the blue end in the $(g-r)_0$ versus CH(G) diagram in Fig.~6.
There are 25 stars (stellar parameters and CH(G) indices are presented in Table 2)
in our sample with carbon abundances available in \cite{Aoki12} 
based on high resolution spectra
for extremely metal poor stars selected from the SDSS survey.
We overplotted the common stars in Fig.~6 according to the carbon to iron
ratio ($[C/Fe]$). It is clear that carbon enhanced metal poor (CEMP) stars with
$[C/Fe]>1.0$ (open diamonds) based on the definition of \cite{Beers05} are located in the upper part while
stars with $[C/Fe]<1.0$ (open triangles) follow our selection regions of RGB stars.
The only exception is the star $2183-53536-175$ (plate-mjd-fiberid)
(with additional open square in Fig.~6), which has $[C/Fe]=1.24$ but the spectra shows a weak CH(G) index.
We suspected a too low $\feh$ was obtained in \cite{Aoki12} and the $[C/Fe]$ reduces to
$0.74$ if the metallicity from SSPP catalog was used. Further study on this
discrepancy is desirable for this star.
Without the carbon abundances in our work, it is impossible to obtain the fraction
of CEMP stars among normal stars. Alternatively,
we selected stars with CH(G) index above $-0.47$ for $(g-r)_0<0.3$
and 0.1 $mag$ higher than the RGB calibration for $0.3\,\leq\,(g-r)_0\,\leq\,0.8$ as being candidates of
high branch CEMP stars, the fraction of which is
16\% for $\feh < -2.8$ and decreases to 5\% at $\feh = -2.0$. This former value is lower than
the fraction of 20\% in \cite{Carollo12} for $\feh < -2.5$, the value of 23\% for $\feh < -3.0$ in \cite{Yong12}
and the value of  28\% in \cite{Norris12} for $\feh < -3.1$.
Note that our selection of high branch CEMP stars may reflect the definition of
$[C/Fe]>1.0$ from \cite{Beers05}, while
\cite{Norris12} adopted the division of $[C/Fe]>0.7$
from \cite{Aoki07}. In addition, our fractions are based on evolved stars with
SSPP $logg < 3.5$, while \cite{Carollo12} include unevolved stars.
In view of the two factors, our value is actually not inconsistent with
those in the literature.

\begin{table*}
\caption{The SDSS identifications, plate-mjd-fiberid numbers, stellar
parameters and $[C/Fe]$ from \cite{Aoki12}, stellar
parameters from SDSS DR9 database and  CH(G) indices from this work
for 25 common stars are presented.}
\label{tb:Mx} \setlength{\tabcolsep}{0.06cm}
\begin{tabular}{rrrrrrrrrr}
\noalign{\smallskip}
\hline
 SDSS ID&plate-mjd-fid&Teff&logg&[Fe/H]&[C/Fe]&Teff&logg&[Fe/H]& CH(G)\\
        &           &HRS  &HRS  &HRS    &HRS & DR9&    Dr9&    Dr9&this\\
\hline
SDSS J0002+2928& 2803-54368-459&6150&4.0&-3.26& 2.63& 6239&3.33& -2.879& -0.414\\
SDSS J0018$-$0939& 1912-53293-352&4600&5.0&-2.65& -0.88& 4601&3.38& -3.012& -0.460\\
SDSS J0126+0607& 2314-53713-090&6900&4.0&-3.01& 3.08& 6799&3.88& -2.860& -0.475\\
SDSS J0259+0057& 1513-53741-338&4550&5.0&-3.31& 0.02& 4578&3.64& -3.704& -0.401\\
SDSS J0308+0505& 2335-53730-314&5950&4.0&-2.19& 2.36& 5961&3.14& -2.423& -0.313\\
SDSS J0351+1026& 2679-54368-543&5450&3.6&-3.18& 1.55& 5631&3.02& -2.773& -0.387\\
SDSS J0711+6702& 2337-53740-564&5350&3.0&-2.91& 1.98& 5396&2.17& -2.914& -0.280\\
SDSS J0723+3637& 2941-54507-222&5150&2.2&-3.32& 1.79& 5076&2.65& -3.480& -0.328\\
SDSS J0741+6708& 2939-54515-414&5200&2.5&-2.87& 0.74& 5252&1.96& -2.965& -0.445\\
SDSS J0858+3541& 2380-53759-094&5200&2.5&-2.53& 0.30& 5150&1.92& -2.885& -0.472\\
SDSS J0912+0216& 0471-51924-613&6150&4.0&-2.68& 2.05& 6208&3.38& -2.545& -0.440\\
SDSS J1036+1212& 1600-53090-378&5850&4.0&-3.47& 1.84& 5847&2.86& -3.307& -0.518\\
SDSS J1241$-$0837& 2689-54149-292&5150&2.5&-2.73& 0.50& 5160&2.57& -2.745& -0.395\\
SDSS J1242$-$0336& 2897-54585-210&5150&2.5&-2.77& 0.64& 5078&2.48& -3.065& -0.444\\
SDSS J1245$-$0738& 2689-54149-491&6100&4.0&-3.17& 2.53& 6212&2.97& -2.894& -0.392\\
SDSS J1349$-$0229& 0913-52433-073&6200&4.0&-3.24& 3.01& 6168&4.39& -3.163& -0.373\\
SDSS J1422+0031& 0304-51609-528&5200&2.2&-3.03& 1.70& 5181&3.03& -3.248& -0.345\\
SDSS J1612+0421& 2178-54629-546&5350&3.3&-2.86& 0.63& 5349&2.49& -3.194& -0.520\\
SDSS J1613+5309& 2176-54243-614&5350&2.1&-3.33& 2.09& 5451&2.70& -2.853& -0.337\\
SDSS J1626+1458& 2202-53566-537&6400&4.0&-2.99& 2.86& 6416&3.38& -2.542& -0.442\\
SDSS J1646+2824& 1690-53475-323&6100&4.0&-3.05& 2.52& 6172&3.20& -2.688& -0.383\\
SDSS J1703+2836& 2808-54524-510&5100&4.8&-3.21& 0.28& 5065&3.48& -3.537& -0.463\\
SDSS J1734+4316& 2799-54368-138&5200&2.7&-2.51& 1.78& 5025&1.98& -3.177& -0.250\\
\noalign{\smallskip}
\hline
\end{tabular}
\end{table*}

\begin{figure*}[ht]
\epsscale{1.0} \plotone{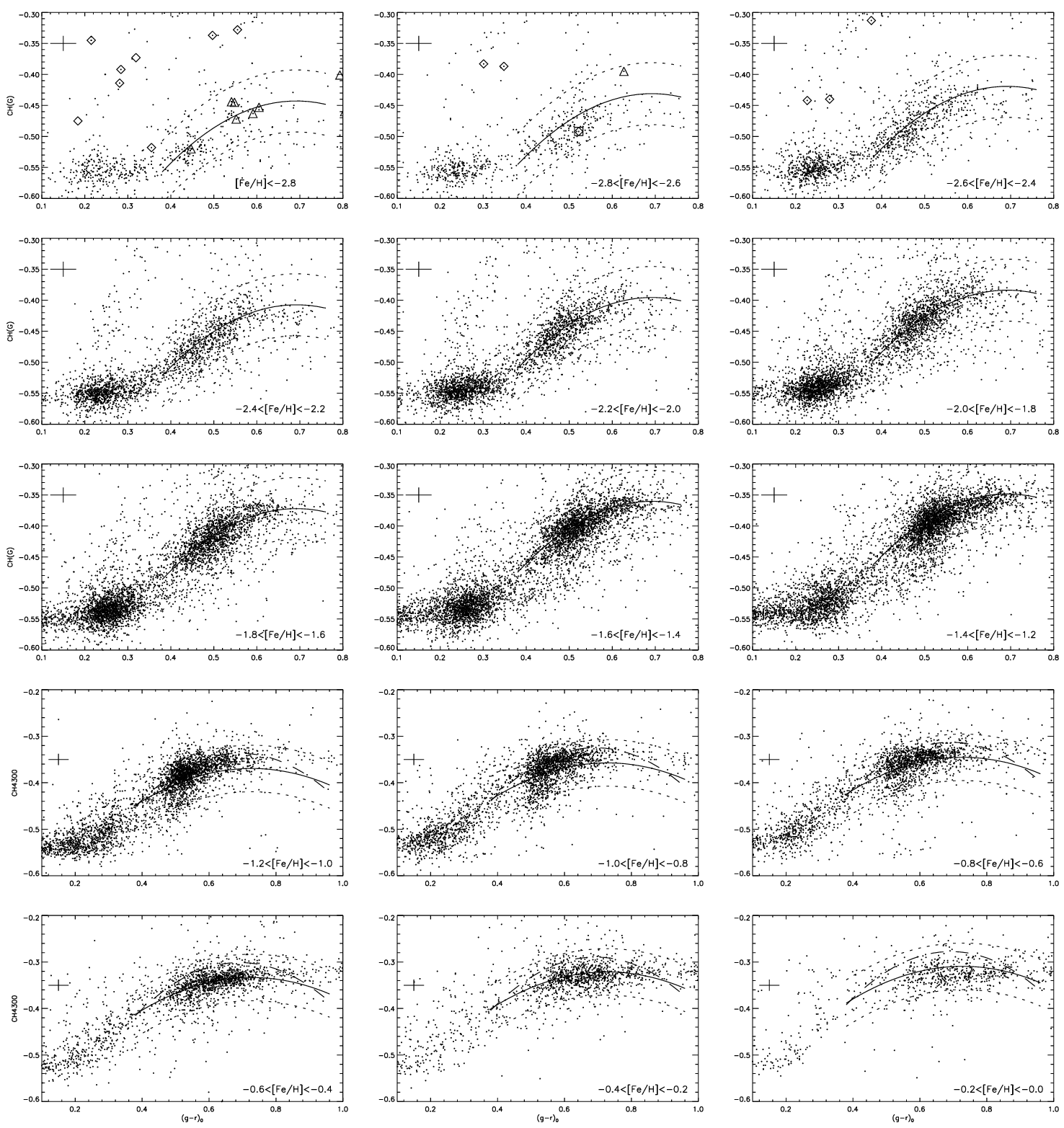} \caption{The $(g-r)_0$ versus CH(G) index
diagram from $\feh < -2.8$ to $\feh = 0.0$ with a bin of 0.2 dex.
The suitable calibration and the selection ranges for RGB stars 
in a given metallicity bin are indicated
by the solid and dashed lines, respectively. For stars with $\feh > -1.2$ the solid lines
are based on the new calibration of $CH(G)=0.953(g-r)_0-0.655(g-r)_0^2+0.060\feh-0.650$, while the long dashed lines
are based on the same calibration of $CH(G)=1.625(g-r)_0-1.174(g-r)_0^2+0.060\feh-0.830$ as 
that of $\feh \leq -1.2$ stars.
Stars with $[C/Fe]>1.0$ from \cite{Aoki12} are indicated by open diamonds and stars with $[C/Fe]<1.0$
are shown as open triangles. The exceptional star with $[C/Fe]=1.24$ is shown by additional open square.
Note that two stars with the CH(G) index lower than $-0.3$ 
are outside of the first panel.}
 \label{fig6}
\end{figure*}

\section{Summary}                         
We have measured the CH(G) index for evolved stars in M3
at $\feh = -1.5$ and found a clear separation between red giant
stars and early-AGB stars at similar colors. The calibration of 
the $(g-r)_0$ versus CH(G) index is established to be
$CH(G) =1.625(g-r)_0-1.174(g-r)_0^2-0.934$
based on field stars selected 
from the SDSS DR9 database with $0.1\,\leq\,(g-r)_0\,\leq\, 0.8$,
$-1.6\,\leq\,\feh\,\leq\,-1.4$ and $logg < 3.5$.
With similar coefficients but a different constant of $-0.982$, the calibration
can be established for $\feh = -2.3$.
Stars selected from the two calibrations within a deviation of
0.05 $mag$ in the CH(G) index at a given $(g-r)_0$ color
 agree well with the locations of RGB
member stars of M3 and M15/M92.
We thus extended this kind
of fitting to other metallicity bins, and the calibration
of $CH(G)=1.625(g-r)_0-1.174(g-r)_0^2+0.060\feh-0.844$ fits well the
location of RGB stars for $-3.0 <\feh \leq -1.2$, but a new calibration of
$CH(G)=0.953(g-r)_0-0.655(g-r)_0^2+0.060\feh-0.650$ for $-1.2 <\feh <0.0$ seems to be more
reasonable
to select RGB stars from evolved stars.
Note that these calibrations are obtained from the SDSS data set and they
may not valid for other data sets because band strength indices are susceptible 
to the overall shape of the spectrum.

We provide an online table for the selected RGB stars.
In the near future, we plan to apply this calibration to select
RGB stars from the SDSS DR9 data set and investigate the
chemical and kinematical properties (mainly based on radial velocities) of the Galaxy. This kind of
stellar tracer can extend to a distance of at least 20 kpc, which
is further than the distance limit of 5 kpc using dwarfs in
the SDSS survey.

\acknowledgments
We thank the referee for providing very good suggestions, which greatly improved
the paper.
This study is supported by the National Natural Science
Foundation of China under grants No. 11233004, 11222326, 11073026, 11150110135, 11078019
and 11103034.

Funding for SDSS-III has been provided by the Alfred P. Sloan Foundation, 
the Participating Institutions, the National Science Foundation, and the U.S. Department 
of Energy Office of Science. The SDSS-III web site is http://www.sdss3.org/.
SDSS-III is managed by the Astrophysical Research Consortium for the Participating 
Institutions of the SDSS-III Collaboration including the University of Arizona, 
the Brazilian Participation Group, Brookhaven National Laboratory, University of 
Cambridge, University of Florida, the French Participation Group, the German 
Participation Group, the Instituto de Astrofisica de Canarias, the Michigan 
State/Notre Dame/JINA Participation Group, Johns Hopkins University, Lawrence 
Berkeley National Laboratory, Max Planck Institute for Astrophysics, New Mexico 
State University, New York University, Ohio State University, Pennsylvania State 
University, University of Portsmouth, Princeton University, the Spanish Participation 
Group, University of Tokyo, University of Utah, Vanderbilt University, University 
of Virginia, University of Washington, and Yale University.


\begin{thebibliography}{}
\bibitem[Abazajian et al. (2009)]{Abazajian09} {Abazajian K.N., Adekman-McCarthy J.K., Agueros M.A., et al. 2009, ApJS 182, 543}
\bibitem[Aihara et al. (2011)]{Aihara11} {Aihara H., Allende Prieto C., An D.,  et al. 2011, ApJS 193, 29} 
\bibitem[Ahn et al. (2012)]{Ahn12} Ahn C., Alexandroff R., Allende Prieto C.,  et al. 2012, ApJS, 203, 21
\bibitem[Allende Prieto et al.(2006)]{Allende06} Allende Prieto, C., Beers, T.~C., Wilhelm, R., et al. 2006, \apj, 636, 804
\bibitem[An et al. (2008)]{An08} An D., Johnson J.A., Clem J.L. et al. 2008, ApJS, 179, 326
\bibitem[Angelou et al. (2011)]{Angelou11} Angelou G.C., Church R.P., Stancliffe R.J., Lattanzio J.C., Smith G.H., 2011, ApJ, 728, 79
\bibitem[Aoki et al. (2007)]{Aoki07} Aoki, W., Beers, T. C., Christlieb, N., Norris, J. E., Ryan, S. G., \& Tsangarides, S. 2007, ApJ, 655, 492
\bibitem[Aoki et al. (2012)]{Aoki12} Aoki, W., Beers, T. C., Lee, Y.S., et al. 2012, AJ, in press (astro-ph/1210.1946)
\bibitem[Beers \& Christlie (2005)]{Beers05} Beers, T. C., \& Christlieb, N. 2005, ARA\&A, 43, 531
\bibitem[Briley et al.(1990)]{Briley90} Briley M.M., Bell R.A., \& Hoban S. 1990, ApJ, 359, 307
\bibitem[Bond et al.(2010)]{Bond10} Bond, N.A., Ivezi\'c, Z.,, Sesar, B. et al. 2010, \apj, 718, 1
\bibitem[Carollo et al.(2010)]{Carollo10} Carollo, D., Beers, T.C., Chiba, M. et al. 2010, \apj, 712, 692
\bibitem[Carollo et al.(2012)]{Carollo12} Carollo, D., Beers T.C., Bovy J. et al. 2012, \apj, 744, 195
\bibitem[Dotter et al. (2008)]{Dotter08} Dotter A., Chaboyer B., Jevremovic D., Kostov V., Baron E., Ferguson J.W., 2008, ApJS, 178, 89
\bibitem[Eisenstein et al. (2011)]{Eisenstein11} Eisenstein D.J., Weinberg D.H., Agol E., et al. 2011, AJ 142, 72 
\bibitem[Harris (1996)]{Harris96} Harris W.E. 1996, AJ, 112, 1487
\bibitem[Lee (1999)]{Lee99} Lee S.G., 1999, AJ, 118, 920
\bibitem[Lee et al. (2008a)]{Lee08a} Lee Y.S., Beers T.C., Sivarani T. et al. 2008, AJ, 136, 2022
\bibitem[Lee et al. (2008b)]{Lee08b} Lee Y.S., Beers T.C., Sivarani T. et al. 2008, AJ, 136, 2050
\bibitem[Norris et al. (1981)]{Norris81} Norris J., Cottrell P.L., Freeman K.C., Da Costa G.S. 1981, ApJ, 244, 205
\bibitem[Norris et al. (2012)]{Norris12} Norris J., Yong D., Bessell M.S. et al. 2012, astro-ph/12113157
\bibitem[Palladino et al. (2012)]{Palladino12} Palladino L.E., Holley-Bockelmann K, Morrison H., et al. 2012, AJ, 143, 128
\bibitem[Smolinski et al. (2011a)]{Smolinski11a} Smolinski J.P., Lee Y.S.,  Beers T.C., et al. 2011, AJ, 141, 89 
\bibitem[Smolinski et al. (2011b)]{Smolinski11b} Smolinski J.P., Martell S.L., Beers T.C., Lee Y.S. 2011, AJ, 142, 126
\bibitem[Suntzeff (1981)]{Suntzeff81} Suntzeff N.B. 1981, ApJS, 47, 1
\bibitem[Sweigart \& Mengel (1979)]{Sweigart79} Sweigart A.V., \& Mengel J.G. 1979, ApJS, 229, 624
\bibitem[Yanny et al. (2009)]{Yanny09} Yanny B., Rockosi C., Newberg H.J., et al. 2009, AJ, 137, 4377
\bibitem[Yong  et al. (2012)]{Yong12} Yong D., Norris J., Bessell M.S. et al. 2012, astro-ph/12083016
\bibitem[Zinn(1973)]{Zinn73} Zinn R., 1973, ApJ, 182, 183
\end{thebibliography}
\end{document}